# Electromagnetic Response for High-Frequency Gravitational Waves in the GHz to THz Band[*]

Fang-Yu Li[†], Meng-Xi Tang[††], and Dong-Ping Shi[†††]

† Department of Physics, Chongqing University, Chongqing 400044, China,

E-mail: cqufangyuli@hotmail.com

†† Department of Physics, Zhongshan University, Guangzhou 510275, China,

E-mail: ststmxt@zsulink.zsu.edu.cn

††† Department of Physics, Chongqing University, Chongqing 400044, China,

E-mail: Shi_dp@hotmail.com

**ABSTRACT**

We consider the electromagnetic (EM) response of a Gaussian beam passing through a static magnetic field to be the high-frequency gravitational waves (HFGW) as generated by several devices discussed at this conference. It is found that under the synchroresonance condition, the first-order perturbative EM power fluxes will contain a "left circular wave" and a "right circular wave" around the symmetrical axis of the Gaussian beam. However, the perturbative effects produced by the states of + polarization and × polarization of the GW have a different physical behavior. For the HFGW of $\nu_g = 3GHz$, $h = 10^{-30}$ (which corresponds to the power flux density $\sim 10^{-6} W \cdot m^{-2}$) to $\nu_g = 1.3 THz$, $h = 10^{-28}$ (which corresponds to the power flux density $\sim 10^3 W \cdot m^{-2}$) expected by the HFGW generators described at this conference, the corresponding perturbative photon fluxes passing through a surface region of $10^{-2} m^2$ would be expected to be $10^3 s^{-1} - 10^4 s^{-1}$. They are the orders of magnitude of the perturbative photon flux we estimated using typical laboratory parameters that could lead to the development of sensitive HFGW receivers. Moreover, we will also discuss the relative background noise problems and the possibility of displaying the HFGW. A laboratory test bed for juxtaposed HFGW generators and our detecting scheme is explored and discussed.

---

I  INTRODUCTION

In recent years, since suggestion of a serried of new theoretical models and physical ideas [1, 2, 3], and fast development of relative technology (such as nanotechnology, ultra-fast science, high-temperature superconductors, ultra-strong laser physics, high energy laboratory astrophysics, etc., [4-9]), these results offered new possibilities for laboratory generation and detection of high-frequency gravitational waves (HFGW).

In this paper we shall study the electromagnetic (EM) response to the HFGW in the GHz-THz band. Our EM detecting system is a Gaussian beam passing through a static magnetic field. Here we consider it since there are following reasons: (1) Unlike the EM response to the gravitational waves (GW's) by an ideal plane EM wave, the Gaussian beam is a realized EM wave beam satisfying physical boundary conditions, and because of the special property of the Gaussian function of the beam, whether the GHz band or the THz region, the EM response has good space accumulation effect. (2) In recent years, strong and ultra-strong lasers and microwave beams have been generated [6-8] under the laboratory conditions, and many of the beams are often expressed as the Gaussian-type or the quasi-Gaussian-type, and they usually have good monochromaticity in the GHz to THz band. (3) The EM response in the GHz to THz band means that the dimensions of the EM detecting system may be reduced to the typical laboratory size (e.g., magnitude of meter). (4) The GHz and THz band are just typical frequency regions expected by the HFGW generators described at this conference (see relevant references in this conference). (5) The Gaussian beam propagating through a static magnetic field is an open system; in this case the EM perturbation might have more direct displaying effect. Therefore, the EM response of the Gaussian beam and other means discussed at this conference have a good complementarity.

II  THE ELECTROMAGNETIC DETECTING SYSTEM: A GAUSSIAN BEAM PASSING THROUGH A STATIC MAGNETIC FIELD.

It is well known that in flat spacetime (i.e., when the GW's are absent) usual form of the fundamental Gaussian beam is given by [10]

$$\psi = \frac{\psi_0}{\sqrt{1+(z/f)^2}} \exp\left(-\frac{r^2}{W^2}\right) \exp\left\{i\left[k_e z - \omega_e t - \tan^{-1}\frac{z}{f} + \frac{k_e r}{2R} + \delta\right]\right\}, \tag{1}$$

where $r^2 = x^2 + y^2$, $k_e = 2\pi/\lambda_e$, $f = \pi W_0^2/\lambda_e$, $W = W_0\left[1+(z/f)^2\right]^{1/2}$, $R = z + f^2/z$, $\psi_0$ is the maximal amplitude of the electric (or magnetic) field of the Gaussian beam, i.e., the amplitude at the plane $z = 0$, $W_0$ is the minimum spot size, namely the spot radius at the plane $z = 0$, $\omega_e$ is the angular frequency of the Gaussian beam, and $\delta$ is an arbitrary phase factor.

Supposing that the electric field of the Gaussian beam is pointed along the direction of the $x$-axis, that it is expressed as Eq. (1), and a static magnetic field pointing along y-axis is localized in the region $-l/2 \leq z \leq l/2$, then we have

$$E^{(0)} = \tilde{E}_x^{(0)} = \psi, \quad E_y^{(0)} = E_z^{(0)} = 0,$$

$$B^{(0)} = \hat{B}^{(0)}, = \begin{cases} \hat{B}_y^{(0)} & (-l/2 \leq z \leq l/2) \\ 0 & (z \leq -l/2 \text{ and } z \geq l/2), \end{cases} \tag{2}$$

where the superscript 0 denotes the background EM fields, the notations $\sim$ and $\wedge$ stand for the time-dependent and static EM fields, respectively. Using (we use MKS units)

$$\tilde{\mathbf{B}}^{(0)} = -\frac{i}{\omega_e} \nabla \times \tilde{\mathbf{E}}^{(0)}, \tag{3}$$

and Eqs. (1) (2), we obtain the time-dependent EM field components in the cylindrical polar coordinates as follows

$$\tilde{E}_r^{(0)} = \psi \cos\phi, \quad \tilde{E}_\phi^{(0)} = -\psi \sin\phi, \quad \tilde{E}_z^{(0)} = 0,$$

$$\tilde{B}_r^{(0)} = -\frac{i}{\omega_e}\frac{\partial \psi}{\partial z}\sin\phi, \quad \tilde{B}_\phi^{(0)} = -\frac{i}{\omega_e}\frac{\partial \psi}{\partial z}\cos\phi, \quad \tilde{B}_z^{(0)} = \frac{i}{\omega_e}\frac{\partial \psi}{\partial y}. \tag{4}$$

For the high-frequency EM power fluxes, only non-vanishing average values of these with respect to time have an observable effect. From Eqs. (1) and (4), one finds

$$<\overset{(0)}{S^z}> = \frac{1}{\mu_0} <\tilde{E}_x^{(0)} \tilde{B}_y^{(0)}>$$

$$= \frac{\psi_0^2}{2\mu_0\omega_e\left[1+(z/f)^2\right]}\left[k_e + \frac{k_e r^2(f^2-z^2)}{2(f^2+z^2)^2} - \frac{f}{f^2+z^2}\right]\exp\left(-\frac{2r^2}{W^2}\right), \tag{5}$$

$$<\overset{(0)}{S^r}> = \frac{1}{\mu_0} <\tilde{E}_\phi^{(0)} \tilde{B}_z^{(0)}>$$
$$= \frac{\psi_0^2 k_e r \sin^2 \phi}{2\mu_0 \omega_e \left[1+(z/f)^2\right](z+f^2/z)} \exp\left(-\frac{2r^2}{W^2}\right), \quad (6)$$

$$<\overset{(0)}{S^\phi}> = -\frac{1}{\mu_0} <\tilde{E}_r^{(0)} \tilde{B}_z^{(0)}>$$
$$= \frac{\psi_0^2 k_e r \sin(2\phi)}{4\mu_0 \omega_e \left[1+(z/f)^2\right](z+f^2/z)} \exp\left(-\frac{2r^2}{W^2}\right), \quad (7)$$

where $<\overset{(0)}{S^z}>$, $<\overset{(0)}{S^r}>$ and $<\overset{(0)}{S^\phi}>$ represent the average values of the axial, radial and tangential EM power flux densities, respectively, the angular brackets denote the average values with respect to time. We can see from Eqs. (5)-(7) that $<\overset{(0)}{S^r}>_{z=0} = <\overset{(0)}{S^\phi}>_{z=0} \equiv 0$, $<\overset{(0)}{S^z}>_{z=0} = <\overset{(0)}{S^z}>_{max}$ and $|<\overset{(0)}{S^z}>| \gg |<\overset{(0)}{S^r}>|$ and $|<\overset{(0)}{S^\phi}>|$ in the region near the minimum spot. Thus the propagation direction of the Gaussian beam is exactly parallel to the z-axis only in the plane $z=0$. In the region of $z \neq 0$, because of nonvanishing $<\overset{(0)}{S^r}>$ and $<\overset{(0)}{S^\phi}>$, the Gaussian beam will be asymptotically spread as $|z|$ increases.

## III THE ELECTROMAGNETIC RESPONSE FOR THE HIGH-FREQUENCY GRAVITATIONAL WAVES OF $\nu_g = 3GHz$ AND $\nu_g = 1.3THz$

### A The perturbation solutions satisfying boundary conditions

The EM response to the weak GW fields can be described by Maxwell equations in curved spacetime, i.e.,

$$\frac{1}{\sqrt{-g}} \frac{\partial}{\partial x^\nu} \left(\sqrt{-g} g^{\mu\alpha} g^{\nu\beta} F_{\alpha\beta}\right) = \mu_0 J^\mu, \quad (8)$$

$$F_{[\mu\nu,\alpha]} = 0, \quad (9)$$

and

$$F_{\mu\nu} = F_{\mu\nu}^{(0)} + \tilde{F}_{\mu\nu}^{(1)}, \quad (10)$$

$$g_{\mu\nu} = \eta_{\mu\nu} + h_{\mu\nu}, \quad (11)$$

where $F^{(0)}_{\mu\nu}$ and $\tilde{F}^{(1)}_{\mu\nu}$ represent the background EM field tensor and the first-order perturbation to $F^{(0)}_{\mu\nu}$ in the presence of the GW, respectively, and $|\tilde{F}^{(1)}_{\mu\nu}| << |F^{(0)}_{\mu\nu}|$ for the nonvanishing $F^{(0)}_{\mu\nu}$ and $\tilde{F}^{(1)}_{\mu\nu}$; $\eta_{\mu\nu}$ is Lorentz metric, $h_{\mu\nu}$ is first-order perturbation to $\eta_{\mu\nu}$ in the presence of the GW. For the EM response in the vacuum, because it has neither the real four-dimensional electric current nor the equivalent electric current caused by the energy dissipation, such as Ohimic losses in the cavity EM response or the dielectric losses, so that $J^{\mu} = 0$ in Eq. (8).

Strictly speaking, the GW's produced by the laboratory generators with finite size are non-plane waves, this property will cause certain difficulty to solve Eqs. (8), (9) and to estimate received the GW energy for the receiver. Fortunately, as we will show that for the HFGW in the GHz to THz band, the above difficulty can be overcome and the HFGW can be seen the quasi-plane waves in the wave zone.

For the HFGW produced by the generators with finite size, the first-order perturbation to flat-spacetime at the optimal radiative direction can be written as

$$h_{\oplus} = h_{xx} = -h_{yy} = \frac{A}{z_0 + z}\exp(ik_{\alpha}x^{\alpha}) = \frac{A}{z_0 + z}\exp[i(k_g z - \omega_g t)],$$

$$h_{\otimes} = h_{xy} = h_{yx} = \frac{iA}{z_0 + z}\exp(ik_{\alpha}x^{\alpha}) = \frac{iA}{z_0 + z}\exp[i(k_g z - \omega_g t)], \quad (12)$$

here we suppose that the optimal radiative direction is pointed along the positive z-axis, $z_0$ is the typical distance between the center of the Gaussian beam and the generators, and the HFGW is circular polarized. Obviously, if the dimensions of the receiver and the wavelength $\lambda_g$ are much less than the distance $z_0$, the small deviation for the propagating direction of the HFGW from the $z$-axis in the receiver region will be negligible. Moreover, if the effective interacting region of the Gaussian beam with the HFGW is localized in $z_0 - l_0 \le z \le z_0 + l_0$ ($l_0 << z_0$), then we have

$$\frac{A}{z_0 - l_0} \approx \frac{A}{z_0 + l_0} \approx \frac{A}{z_0}. \quad (13)$$

From Eq. (12), the first- and second-order derivations to z in Eqs.(8) and (9) can be given by

$$\frac{\partial h_{\oplus}}{\partial z} = -\frac{A}{(z_0 + z)^2}\exp(ik_{\alpha}x^{\alpha}) + \frac{ik_g A}{z_0 + z}\exp[i(k_{\alpha}x^{\alpha})],$$

$$\frac{\partial^2 h_{\oplus}}{\partial z^2} = -\frac{2A}{(z_0 + z)^3}\exp(ik_{\alpha}x^{\alpha}) - \frac{i2k_g A}{(z_0 + z)^2}\exp[i(k_{\alpha}x^{\alpha})] - \frac{k_g^2 A}{z_0 + z}\exp[i(k_{\alpha}x^{\alpha})]. \quad (14)$$

Setting $z_0 = 3m$, $l_0 = 0.3m$, and $\nu_g = 3GHz$, one finds

$$\left|\frac{2A}{(z_0+z)^3}\right|:\left|\frac{2k_gA}{(z_0+z)^2}\right|:\left|\frac{k_g^2A}{z_0+z}\right|\approx 1:1.99\times 10^2:1.88\times 10^4, \tag{15}$$

putting $z_0=3m$, $l_0=0.3m$, and $\nu_g=1.3THz$, we have

$$\left|\frac{2A}{(z_0+z)^3}\right|:\left|\frac{2k_gA}{(z_0+z)^2}\right|:\left|\frac{k_g^2A}{z_0+z}\right|\approx 1:8.71\times 10^4:3.54\times 10^9. \tag{16}$$

Clearly, the state of $\times$ polarization $h_\otimes$ of the HFGW has the same property as Eqs.(15) and (16). Therefore, whether the 3GHz HFGW or 1.3THz HFGW, we have [see Eqs.(14)-(16)]

$$\begin{aligned}\left|\frac{\partial h_\oplus}{\partial z}\right|=\left|\frac{\partial h_\otimes}{\partial z}\right|&\approx\frac{k_gA}{z_0+z}|\exp(ik_\alpha x^\alpha)|\\&=\frac{A}{z_0+z}\left|\frac{\partial}{\partial z}\exp(ik_\alpha x^\alpha)\right|=A_\oplus\left|\frac{\partial}{\partial z}\exp(ik_\alpha x^\alpha)\right|,\\\left|\frac{\partial^2 h_\oplus}{\partial z^2}\right|=\left|\frac{\partial^2 h_\otimes}{\partial z^2}\right|&\approx\frac{k_g^2A}{z_0+z}|\exp(ik_\alpha x^\alpha)|\\&=\frac{A}{z_0+z}\left|\frac{\partial^2}{\partial z^2}[\exp(ik_\alpha x^\alpha)]\right|=A_\oplus\left|\frac{\partial^2}{\partial z^2}\exp(ik_\alpha x^\alpha)\right|,\end{aligned} \tag{17}$$

where

$$A_\oplus=A_\otimes=\frac{A}{z_0+z}\approx\frac{A}{z_0},\qquad (z_0-l_0\le z\le z_0+l_0,\ l_0<<z_0). \tag{18}$$

Eq. (17) shows that the HFGW in the effectively receiver region can be treated the quasi-plane GW's. Under the circumstances, Eq. (12) can be rewritten as

$$\begin{aligned}h_\oplus=h_{xx}=-h_{yy}&=A_\oplus\exp[i(k_gz-\omega_g t)],\\h_\otimes=h_{xy}=h_{yx}&=iA_\otimes\exp[i(k_gz-\omega_g t)].\end{aligned} \tag{19}$$

This is just usual form of the GW in the TT gauge. Eq. (19) can be viewed as the approximation expression of the plane wave.

In our EM system, in order to find the first-order perturbation $\tilde{F}^{(1)}_{\mu\nu}$, it is necessary to solve Eqs. (8) and (9) by substituting Eqs.(1), (2), (4) and (19) into them, this is often also quite difficult, even if Eq. (12) and (19) can be seen the form of the plane wave. However, as shown in Refs. [11, 12], the amplitude ratio of the first-order perturbation EM fields produced by the direct interaction of the GW with the EM wave (Gaussian beam) and the static magnetic field is approximately $h\tilde{B}^{(0)}/h\hat{B}^{(0)}$. In our case, we have chosen $\tilde{B}^{(0)}\le 10^{-3}T$, $\hat{B}^{(0)}\sim 10T$, i.e., their ratio is only $10^{-4}$ or less. Thus the former can be neglected. In other words, the contribution of the Gaussian beam is mainly expressed as the coherent synchroresonance (i.e., $\omega_e = \omega_g$) of it with the first-order perturbation $\tilde{F}^{(1)}_{\mu\nu}$ generated by the direct interaction of the HFGW with the static field $\hat{B}^{(0)}_y$. In this case, the process of solving Eqs. (8) and (9) can be greatly simplified. Under these circumstances, we obtain the pure perturbative EM fields [the real part of the perturbation solutions of Eqs. (8) and (9)] in the laboratory frame of references as follows [13]:

(a) Region I ($z \leq -l/2$, $\hat{B}^{(0)} = 0$):

$$\tilde{E}^{(1)}_x = c\tilde{F}^{(1)}_{01} = \tilde{E}^{(1)}_y = c\tilde{F}^{(1)}_{02} = 0,$$

$$\tilde{B}^{(1)}_x = \tilde{F}^{(1)}_{32} = \tilde{B}^{(1)}_y = \tilde{F}^{(1)}_{13} = 0. \tag{20}$$

(b) Region II ($-l/2 \leq z \leq l/2$, $\hat{B}^{(0)} = \hat{B}^{(0)}_y$):

$$\tilde{E}^{(1)}_x = -\frac{1}{2} A_\oplus \hat{B}^{(0)}_y k_g c(z+l/2)\sin(k_g z - \omega_g t) - \frac{1}{2} A_\oplus \hat{B}^{(0)}_y c \sin(k_g z)\sin(\omega_g t),$$

$$\tilde{B}^{(1)}_y = -\frac{1}{2} A_\oplus \hat{B}^{(0)}_y k_g (z+l/2)\sin(k_g z - \omega_g t) + \frac{1}{2} A_\oplus \hat{B}^{(0)}_y \sin(k_g z)\sin(\omega_g t), \tag{21}$$

$$\tilde{E}^{(1)}_y = -\frac{1}{2} A_\otimes \hat{B}^{(0)}_y k_g c(z+l/2)\cos(k_g z - \omega_g t) - \frac{1}{2} A_\otimes \hat{B}^{(0)}_y c \sin(k_g z)\cos(\omega_g t),$$

$$\tilde{B}^{(1)}_x = \frac{1}{2} A_\otimes \hat{B}^{(0)}_y k_g (z+l/2)\cos(k_g z - \omega_g t) - \frac{1}{2} A_\otimes \hat{B}^{(0)}_y \sin(k_g z)\cos(\omega_g t), \tag{22}$$

(c) Region III ($l/2 \leq z \leq l_0$, $\hat{B}^{(0)} = 0$):

$$\tilde{E}^{(1)}_x = -\frac{1}{2} A_\oplus \hat{B}^{(0)}_y k_g cl \sin(k_g z - \omega_g t),$$

$$\tilde{B}^{(1)}_y = -\frac{1}{2} A_\oplus \hat{B}^{(0)}_y k_g l \sin(k_g z - \omega_g t), \tag{23}$$

$$\tilde{E}^{(1)}_y = -\frac{1}{2} A_\otimes \hat{B}^{(0)}_y k_g cl \cos(k_g z - \omega_g t),$$

$$\tilde{B}^{(1)}_x = \frac{1}{2} A_\otimes \hat{B}^{(0)}_y k_g l \cos(k_g z - \omega_g t). \tag{24}$$

and

$$l = n\lambda_g \quad (n \text{ is integer}), \tag{25}$$

where $l_0$ is also the size of the effective region in which the second-order perturbative EM power fluxes, such as $\frac{1}{\mu_0}(\tilde{E}_x^{(1)}\tilde{B}_y^{(1)})$, $\frac{1}{\mu_0}(\tilde{E}_y^{(1)}\tilde{B}_x^{(1)})$, keep a plane wave form. It is easily to show that the perturbative EM fields, Eqs.(20)-(24), satisfy the boundary conditions (the continuity conditions):

$$\left(\tilde{F}_{\mu\nu I}^{(1)}\right)_{z=-l/2} = \left(\tilde{F}_{\mu\nu II}^{(1)}\right)_{z=-l/2}, \quad \left(\tilde{F}_{\mu\nu II}^{(1)}\right)_{z=l/2} = \left(\tilde{F}_{\mu\nu III}^{(1)}\right)_{z=l/2}. \tag{26}$$

Since the weakness of the interaction of the GW's with the EM fields, we shall focus our attention on the first-order perturbative power fluxes produced by the coherent synchroresonance of the above perturbative EM fields with the background Gaussian beam.

## B  The first-order perturbative EM power fluxes

In the following we shall consider only the first-order tangential perturbative power flux density

$$\overset{(1)}{S^\phi} = -\frac{1}{\mu_0}(\tilde{E}_r^{(1)}\tilde{B}_z^{(0)}) = -\frac{1}{\mu_0}(\tilde{E}_x^{(1)}\tilde{B}_z^{(0)})\cos\phi - \frac{1}{\mu_0}(\tilde{E}_y^{(1)}\tilde{B}_z^{(0)})\sin\phi. \tag{27}$$

As we have pointed out above, that for the high-frequency perturbative power fluxes, only nonvanishing average values of them with respect to time have observable effect. It is easily seen from Eqs.(1), (4) and (21)-(24), that average values of Eq.(27) with respect to time, vanish in whole frequency range of $\omega_e \neq \omega_g$, In other words, only under the synchroresonance condition of $\omega_e = \omega_g$, $\overset{(1)}{S^\phi}$ has nonvanishing average value with respect to time. Introducing Eqs.(1), (4) and (21)-(24) into (27), and setting $\delta = \pi/2$, in Eq.(1) (this is always possible), we obtain

$$\left\langle \overset{(1)}{S^\phi} \right\rangle_{\omega_e=\omega_g} = \left\langle \overset{(1)}{S^\phi_\oplus} \right\rangle_{\omega_e=\omega_g} + \left\langle \overset{(1)}{S^\phi_\otimes} \right\rangle_{\omega_e=\omega_g}, \tag{28}$$

where

$$\left\langle \overset{(1)}{S^\phi_\oplus} \right\rangle_{\omega_e=\omega_g} = -\frac{1}{\mu_0}\left\langle \tilde{E}_x^{(1)}\tilde{B}_z^{(0)} \right\rangle\cos\phi, \tag{29}$$

$$\left\langle \overset{(1)}{S^\phi_\otimes} \right\rangle_{\omega_e=\omega_g} = -\frac{1}{\mu_0}\left\langle \tilde{E}_y^{(1)}\tilde{B}_z^{(0)} \right\rangle\sin\phi. \tag{30}$$

$\left\langle \overset{(1)}{S^\phi_\oplus} \right\rangle_{\omega_e=\omega_g}$ and $\left\langle \overset{(1)}{S^\phi_\otimes} \right\rangle_{\omega_e=\omega_g}$ represent the average values of the first-order tangential perturbative power flux density generated by the states of $+$ polarization and $\times$ polarization of the HFGW, Eq.(19), respectively. Using Eqs.(1), (4), (20)-(24) and the boundary conditions, Eqs. (25) and (26), one finds

(a) Region I ($z \leq l/2$),

$$\left\langle \overset{(1)}{S^\phi_\oplus} \right\rangle_{\omega_e = \omega_g} = \left\langle \overset{(1)}{S^\phi_\otimes} \right\rangle_{\omega_e = \omega_g} = 0. \tag{31}$$

(b) Region II ($-l/2 \leq z \leq l/2$),

$$\left\langle \overset{(1)}{S^\phi_\oplus} \right\rangle_{\omega_e = \omega_g} = \left\{ \frac{A_\oplus \hat{B}^{(0)}_y \psi_0 k_g r(z+l/2)}{8\mu_0 [1+(z/f)^2]^{1/2}(z+f^2/z)} \cos\left(\tan^{-1}\frac{z}{f} - \frac{k_g r^2}{2R}\right) \right.$$
$$+ \frac{A_\oplus \hat{B}^{(0)}_y \psi_0 r(z+l/2)}{4\mu_0 W_0^2 [1+(z/f)^2]^{3/2}} \sin\left(\tan^{-1}\frac{z}{f} - \frac{k_g r^2}{2R}\right)$$
$$- \frac{A_\oplus \hat{B}^{(0)}_y \psi_0 r}{8\mu_0 [1+(z/f)^2]^{1/2}(z+f^2/z)} \sin(k_g z) \cos\left(\tan^{-1}\frac{z}{f} - \frac{k_g r^2}{2R} - k_g z\right) \tag{32}$$
$$\left. - \frac{A_\oplus \hat{B}^{(0)}_y \psi_0 r}{4\mu_0 k_g W_0^2 [1+(z/f)^2]^{3/2}} \sin(k_g z) \sin\left(\tan^{-1}\frac{z}{f} - \frac{k_g r^2}{2R} - k_g z\right) \right\}$$
$$\times \exp(-\frac{r^2}{W^2}) \sin(2\phi),$$

$$\left\langle \overset{(1)}{S^\phi_\otimes} \right\rangle_{\omega_e = \omega_g} = \left\{ \frac{A_\otimes \hat{B}^{(0)}_y \psi_0 k_g r(z+l/2)}{4\mu_0 [1+(z/f)^2]^{1/2}(z+f^2/z)} \sin\left(\frac{k_g r^2}{2R} - \tan^{-1}\frac{z}{f}\right) \right.$$
$$+ \frac{A_\otimes \hat{B}^{(0)}_y \psi_0 r(z+l/2)}{2\mu_0 W_0^2 [1+(z/f)^2]^{3/2}} \cos\left(\frac{k_g r^2}{2R} - \tan^{-1}\frac{z}{f}\right)$$
$$+ \frac{A_\otimes \hat{B}^{(0)}_y \psi_0 r}{4\mu_0 [1+(z/f)^2]^{1/2}(z+f^2/z)} \sin(k_g z) \sin\left(k_g z - \tan^{-1}\frac{z}{f} + \frac{k_g r^2}{2R}\right) \tag{33}$$
$$\left. + \frac{A_\otimes \hat{B}^{(0)}_y \psi_0 r}{2\mu_0 k_g W_0^2 [1+(z/f)^2]^{3/2}} \sin(k_g z) \cos\left(k_g z - \tan^{-1}\frac{z}{f} + \frac{k_g r^2}{2R}\right) \right\}$$
$$\times \exp(-\frac{r^2}{W^2}) \sin^2 \phi.$$

(c) Region III ($l/2 \leq z \leq l_0$),

$$\left\langle \overset{(1)}{S^\phi_\oplus} \right\rangle_{\omega_e = \omega_g} = \left\{ \frac{A_\oplus \hat{B}^{(0)}_y \psi_0 k_g lr}{8\mu_0 [1+(z/f)^2]^{1/2}(z+f^2/z)} \cos\left(\tan^{-1}\frac{z}{f} - \frac{k_g r^2}{2R}\right) \right.$$
$$\left. + \frac{A_\oplus \hat{B}^{(0)}_y \psi_0 lr}{4\mu_0 W_0^2 [1+(z/f)^2]^{3/2}} \sin\left(\tan^{-1}\frac{z}{f} - \frac{k_g r^2}{2R}\right) \right\} \tag{34}$$
$$\times \exp(-\frac{r^2}{W^2}) \sin(2\phi),$$

$$\langle S_{\otimes}^{\phi}\rangle_{\omega_e=\omega_g}^{(1)} = \left\{ \frac{A_{\otimes}\hat{B}_y^{(0)}\psi_0 k_g l}{4\mu_0[1+(z/f)^2]^{1/2}(z+f^2/z)}\sin\left(\frac{k_g r^2}{2R} - \tan^{-1}\frac{z}{f}\right) \right.$$

$$\left. + \frac{A_{\otimes}\hat{B}_y^{(0)}\psi_0 lr}{2\mu_0 W_0^2[1+(z/f)^2]^{3/2}}\cos\left(\frac{k_g r^2}{2R} - \tan^{-1}\frac{z}{f}\right) \right\} \exp(-\frac{r^2}{W^2})\sin^2\phi. \quad (35)$$

Eqs. (32)-(35) show that because there are nonvanishing $\langle S_{\oplus}^{\phi}\rangle_{\omega_e=\omega_g}^{(1)}$ (which depends on the $+$ polarization state of the HFGW) and $\langle S_{\otimes}^{\phi}\rangle_{\omega_e=\omega_g}^{(1)}$ (which depends on the $\times$ polarization state of the HFGW), the first-order tangential perturbative power fluxes are expressed as the "left circular wave" and "right circular wave" in the cylindrical polar coordinates, which around the symmetrical axis of the Gaussian beam, but $\langle S_{\oplus}^{\phi}\rangle_{\omega_e=\omega_g}^{(1)}$ and $\langle S_{\otimes}^{\phi}\rangle_{\omega_e=\omega_g}^{(1)}$ have a different physical behavior. By comparing Eqs. (32)-(35) with Eqs. (5)-(7), we can see that: (a) $\langle S_{\oplus}^{\phi}\rangle_{\omega_e=\omega_g}^{(1)}$ and $\langle S^{\phi}\rangle^{(0)}$ have the same angular distribution factor $\sin(2\phi)$, thus $\langle S_{\oplus}^{\phi}\rangle_{\omega_e=\omega_g}^{(1)}$ will be swamped by the background power flux $\langle S^{\phi}\rangle^{(0)}$. Namely, in this case $\langle S_{\oplus}^{\phi}\rangle_{\omega_e=\omega_g}^{(1)}$ has no observable effect. (b) The angular distribution factor of $\langle S_{\otimes}^{\phi}\rangle_{\omega_e=\omega_g}^{(1)}$ is $\sin^2\phi$, which is different from that of $\langle S^{\phi}\rangle^{(0)}$. Therefore, $\langle S_{\otimes}^{\phi}\rangle_{\omega_e=\omega_g}^{(1)}$, in principle, has an observable effect. In particular, at the surfaces $\phi = \pi/2$ and $3\pi/2$, $\langle S^{\phi}\rangle^{(0)} \equiv 0$, while $|\langle S_{\otimes}^{\phi}\rangle_{\omega_e=\omega_g}^{(1)}| = |\langle S_{\otimes}^{\phi}\rangle_{\omega_e=\omega_g}^{(1)}|_{\max}$, this is satisfactory (although $\langle S^r\rangle^{(0)}$ and $\langle S_{\otimes}^{\phi}\rangle_{\omega_e=\omega_g}^{(1)}$ have the same angular distribution factor $\sin^2\phi$, the propagating direction of $\langle S^r\rangle^{(0)}$ is perpendicular to that of $\langle S_{\otimes}^{\phi}\rangle_{\omega_e=\omega_g}^{(1)}$, thus $\langle S^r\rangle^{(0)}$ (including $\langle S^z\rangle^{(0)}$) has no contribution in the pure tangential direction).

Figure 1 gives the distribution of $\langle S_{\oplus}^{\phi}\rangle_{\omega_e=\omega_g}^{(1)}$ at the plane $z = l/2 = \frac{n}{2}\lambda_g$ ($n$ is integer) in the cylindrical polar coordinates. Figure 2 gives the distribution of $\langle S_{\otimes}^{\phi}\rangle_{\omega_e=\omega_g}^{(1)}$ at the plane $z = 0$ in the cylindrical polar coordinates.

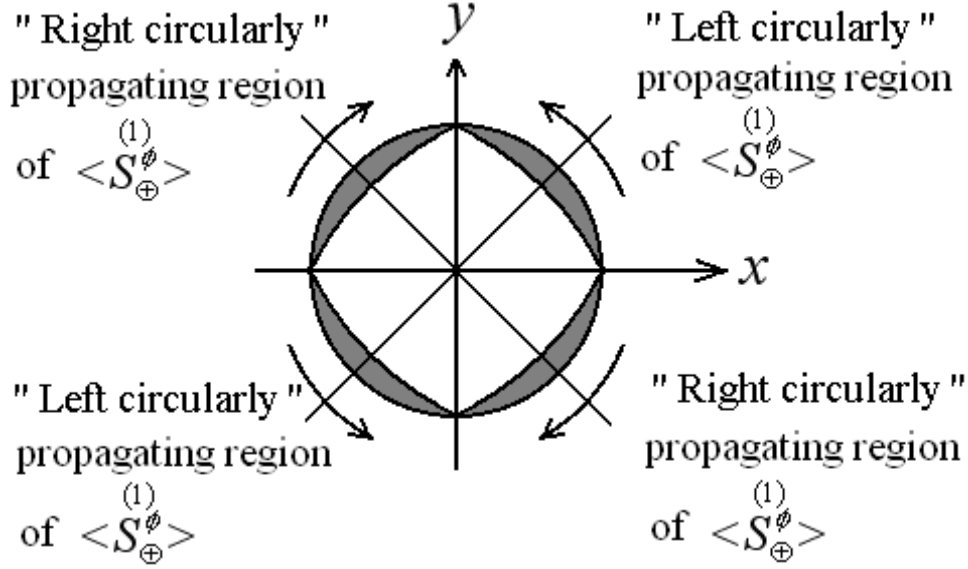

Figure. 1   Distribution of $\langle S_{\oplus}^{\phi} \rangle^{(1)}_{\omega_e=\omega_g}$ at the plane $z = l/2 = n/2\lambda_g$ ($n$ is integer) in the cylindrical polar coordinates. It has maxima at $\phi = \pi/4, 3\pi/4, 5\pi/4$ *and* $7\pi/4$, while it vanishes at $\phi = 0, \pi/2, \pi$ *and* $3\pi/2$. Here $l = 0.1m$ and $\lambda_g = r = 0.1m$, and the GW propagates along the $z$ axis.

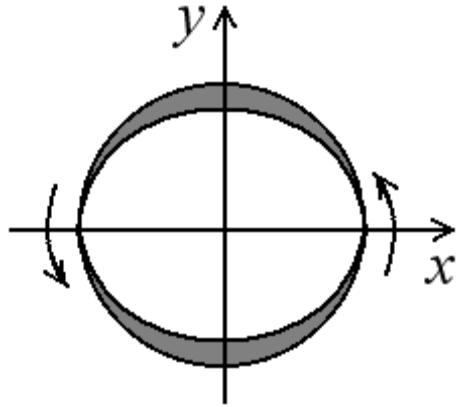

Figure. 2   Distribution of $\langle S_{\otimes}^{\phi} \rangle^{(1)}_{\omega_e=\omega_g}$ at the plane $z = 0$ in the cylindrical polar coordinates. It has maxima at $\phi = \pi/2$ *and* $3\pi/2$, while it vanishes at $\phi = 0, \pi$. Unlike Figure 1, $\langle S_{\otimes}^{\phi} \rangle^{(1)}_{\omega_e=\omega_g}$ at the plane $z = 0$ is completely "left-hand circular". Here $\lambda_g = r = 0.1m$, and the GW propagates along the $z$ axis.

Figures 1, 2 and Eqs. (31)-(35) show that the planes $\phi = \pi/2$ and $3\pi/2$ are the two most interesting regions, in which $\langle \overset{(0)}{S^\phi} \rangle = \langle \overset{(1)}{S^\phi_\oplus} \rangle_{\omega_e=\omega_g} \equiv 0$, but where $|\langle \overset{(1)}{S^\phi_\otimes} \rangle_{\omega_e=\omega_g}| = |\langle \overset{(1)}{S^\phi_\otimes} \rangle_{\omega_e=\omega_g}|_{\max}$. This means that any nonvanishing tangential EM power flux passing through the above surfaces will express a pure EM perturbation produced by the HFGW.

### C  Numerical estimations

If we describe the perturbation in the quantum language (photon flux), the corresponding perturbative photon flux $n_\phi$ caused by $\langle \overset{(1)}{S^\phi_\otimes} \rangle_{\omega_e=\omega_g}$ at the surface $\phi = \pi/2$ (we note that $\langle \overset{(1)}{S^\phi_\otimes} \rangle_{\omega_e=\omega_g}$ is the unique nonvanishing power flux density passing through the surface) can be given by

$$n_\phi = \frac{\langle \overset{(1)}{u^\phi_\otimes} \rangle_{\omega_e=\omega_g}}{\hbar \omega_e} = \frac{1}{\hbar \omega_e} \int_0^{w_0} \int_{-l/2}^{l_0} \langle \overset{(1)}{S^\phi_\otimes} \rangle_{\omega_e=\omega_g, \phi=\pi/2} \, dz dr, \tag{36}$$

where $\langle \overset{(1)}{u^\phi_\otimes} \rangle_{\omega_e=\omega_g} = \int_0^{w_0} \int_{-l/2}^{l_0} \langle \overset{(1)}{S^\phi_\otimes} \rangle_{\omega_e=\omega_g, \phi=\pi/2} \, dz dr$ is the perturbative power flux passing through the surface $\phi = \pi/2$, $\hbar$ is the Planck constant.

#### 1.  The electromagnetic response to the 3GH HFGW

In order to give reasonable estimation, we choose the achievable values of the EM parameters in the present experiments: (1) $P = 10^5 W$, $10^3 W$ and $10W$, respectively, the power of the Gaussian beam. If the spot radius $W_0$ of the Gaussian beam is limited in $0.1m$, then the power can be estimated as $P = \int_0^{W_0} \langle \overset{(0)}{S^z} \rangle_{z=0} 2\pi r dr$ [see Eq. (5)], the values of $\psi_0$ will be $1.31 \times 10^5 Vm^{-1}$, $1.31 \times 10^4 Vm^{-1}$ and $1.31 \times 10^3 Vm^{-1}$, respectively. The above power is well within the current technology condition [6-8]. (2) $\hat{B}_y^{(0)} = 30T$, the strength of the background static magnetic field, this is also achievable strength of a stationary magnetic field under the present experimental condition [14]. (3) $A_\otimes = 10^{-30}$, $\nu_g = 3GHz$. Substituting Eqs. (33), (35) and the above parameters into Eq. (36), and setting $W_0 = l = 0.1m$, $l_0 = 0.3m$, we obtain $n_\phi = 4.75 \times 10^3 s^{-1}$, $4.75 \times 10^2 s^{-1}$ and $4.75 \times 10 s^{-1}$, respectively (see Table I). If the integration region of the radial coordinate $r$ in Eq. (36) is moved to $W_0 \leq r \leq r_0$ (where $W_0 = 0.1m$, $r_0 = 0.2m$), in the same way, the corresponding perturbative photon flux $n'_\phi$ can be estimated as

$$n'_\phi = \frac{1}{\hbar \omega_e} \int_{W_0}^{r_0} \int_{-l/2}^{l_0} \langle S_\otimes^{\phi (1)} \rangle_{\omega_e = \omega_g, \phi = \pi/2} dz dr \approx 4.46 \times 10^3 s^{-1}, \ 4.46 \times 10^2 s^{-1} \text{ and } 4.46 \times 10 s^{-1}, \text{ respectively.}$$

We can see that $n'_\phi < n_\phi$, but $n'_\phi$ retains basically the same order of magnitude as $n_\phi$, and because then the "receiving surface" of the tangential photon flux has already moved to the region outside the spot radius $W_0$ of the Gaussian beam, this result has more realistic meaning to distinguish and display the perturbative photon flux.

Table I. The perturbative photon fluxes $n_\phi$, $n'_\phi$, the power of the background Gaussian beam and corresponding relevant parameters.

|     | $A$ | $\nu_g (Hz)$ | $P(W)$ | $\langle u_\otimes^{\phi (1)} \rangle_{\omega_e = \omega_g} (W)$ | $n_\phi (s^{-1})$ | $n'_\phi (s^{-1})$ |
|-----|-----|------|-----|------|------|------|
| (1) | $10^{-30}$ | $3 \times 10^9$ | $10^5$ | $0.95 \times 10^{-20}$ | $4.75 \times 10^3$ | $4.46 \times 10^3$ |
| (2) | $10^{-30}$ | $3 \times 10^9$ | $10^3$ | $0.95 \times 10^{-21}$ | $4.75 \times 10^2$ | $4.46 \times 10^2$ |
| (3) | $10^{-30}$ | $3 \times 10^9$ | $10$ | $0.95 \times 10^{-22}$ | $4.75 \times 10$ | $4.46 \times 10$ |

We emphasize that here $n_\phi$, $n'_\phi \propto \psi_0 \propto \sqrt{P}$ [see Eqs. (33)(35) and (36)], and at the same time, $n_\phi$, $n'_\phi \propto \hat{B}_y^{(0)}$. Therefore, even if $P$ is reduced to $10W$ (this is already a very relaxed requirement), we have still $n_\phi \approx 4.75 \times 10 s^{-1}$, $n'_\phi \approx 4.46 \times 10 s^{-1}$ (see Table I). Thus, if possible, increasing $\hat{B}_y^{(0)}$ (in this way the number of the background real photons does not change) has better physical effect than increasing $P$.

2. **The electromagnetic response to the** $1.3THz$ **HFGW**

According to the above discussions, it is easily to study the EM response to the $1.3THz$ HFGW. Obviously, in this case the synchroresonance condition $\omega_e = \omega_g$ can be directly satisfied as long as the frequency of the Gaussian beam is tuned to the $1.3THz$.

In the same way, we list the power $P$ of the Gaussian beam, the perturbative power flux passing through the surface $\phi = \pi/2$ ($\sim 10^{-2} m^2$), the perturbative photon fluxes $n_\phi$, $n'_\phi$ and the other relevant parameters in Table II.

Table II. The perturbative photon fluxes $n_\phi$, $n'_\phi$, the power of the background Gaussian beam and corresponding relevant parameters.

|     | $A$ | $\nu_g(Hz)$ | $P(W)$ | $\langle u_\otimes^\phi \rangle^{(1)}_{\omega_e=\omega_g}(W)$ | $n_\phi(s^{-1})$ | $n'_\phi(s^{-1})$ |
| --- | --- | --- | --- | --- | --- | --- |
| (1) | $10^{-28}$ | $1.3 \times 10^{12}$ | $10^5$ | $1.46 \times 10^{-18}$ | $1.69 \times 10^3$ | $0.94 \times 10^3$ |
| (2) | $10^{-28}$ | $1.3 \times 10^{12}$ | $10^3$ | $1.46 \times 10^{-19}$ | $1.69 \times 10^2$ | $0.94 \times 10^2$ |
| (3) | $10^{-28}$ | $1.3 \times 10^{12}$ | $10$ | $1.46 \times 10^{-20}$ | $1.69 \times 10$ | $0.94 \times 10$ |

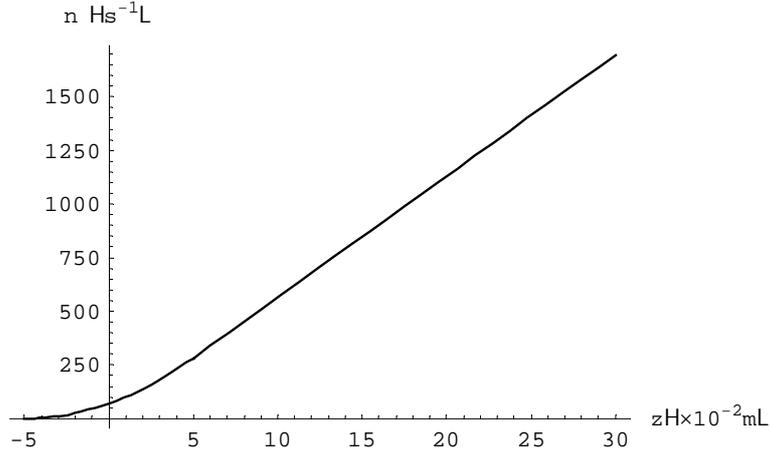

FIG 3 Rating curve between $n_\phi$ and the axis coordinate $z$, here $\nu_e = \nu_g = 1.3 \times 10^{12} Hz$, $A_\otimes = 10^{-28}$, $\hat{B}_y^{(0)} = 30T$, $\psi_0 = 1.29 \times 10^5 Vm^{-1}$, and $W_0 = 0.1m$. It shows that the tangential perturbative photon flux passing through the surface $\phi = \pi/2$ ($\sim 10^{-2} m^2$), would be expected to be $1.69 \times 10^3 s^{-1}$.

Figure 3 gives the rating curve between $n_\phi$ and the axis coordinate $z$, and relative parameters are chosen as $\nu_e = \nu_g = 1.3 THz$, $A_\otimes = 10^{-28}$, $\hat{B}_y^{(0)} = 30T$, $\psi_0 = 1.29 \times 10^5 Vm^{-1}$, $l = 0.1m$, $l_0 = 0.3m$, and $W_0 = 0.1m$. Figure 3 shows that $n_\phi$ has a good accumulation effect as $z$ increases.

## IV. NOISE ISSUES

Since in our EM system the response frequencies are much higher than ones of usual environment noise sources (e.g., mechanical, seismic noise, etc.), the requirement of relevant conditions can be further relaxed. Moreover, this EM system has the following advantages to reduce noise: (1) For the possible external EM noise sources, the use of a Faraday cage or special fractal materials would be very helpful. Once the EM system is isolated from the outside world by such means, the possible noise sources would be the remaining thermal photons and self-background action. (2) Because of the "random motion" of the remaining thermal photons and the highly directional propagated property of the perturbative photon fluxes, the influence of such noise would be effectively suppressed in the local regions. (3) Since the background photon fluxes vanish while the perturbative photon fluxes have maxima in some special regions (e.g., the surfaces $\phi = \pi/2$ and $\phi = 3\pi/2$), such that the signal-to-noise ratio would be much larger than unity at these surfaces, although the background photon fluxes are much greater than the perturbative photon fluxes in other regions. (4) Even if for the mixed regions of the signal and the background action, because the self-background action would decay as $\exp\left(-\frac{2r^2}{W^2}\right)$ [see Eqs. (5)-(7)] while the signal (the perturbative photon fluxes) would decay as $\exp\left(-\frac{r^2}{W^2}\right)$ [see Eqs. (34) and (35)], the signal-to-noise ratio may reach up to unity in the regions of $r = 3W$ or $4W$, roughly. Of course, in such regions the signal will be reduced to a very small value, but increasing background static field (if possible) may be a better way, because in this way the number of the background real photons does not change. Besides, low-temperature vacuum operation might effectively reduce the frequency of the remaining thermal photons and avoid dielectric dissipation. Therefore, the crucial parameters for the noise problems are the selected perturbative photon fluxes passing through the special surfaces and not the all background photons.

## V  A POSSIBLE JUXTAPOSED TEST SCHEME FOR THE ELECTROMAGNETIC DETECTING SYSTEM AND THE HFGW GENERATORS

We shall consider the juxtaposed test scheme for our EM detecting system and 3GHz HFGW and 1.3THz HFGW generators.

Supposing that the EM detecting system is just laid at the optimal radiative direction, and the HFGW in the wave zone can be seen as a quasi-plane GW (as we have pointed out in section III), then the average power flux density of the HFGW of the $\times$ polarization state in the wave zone can be estimated as

$$\langle S_{GW} \rangle \approx \frac{c^3 \omega^2}{32\pi G} A_\otimes^2. \tag{37}$$

Thus the total power flux of the HFGW passing through the effective interaction region (cross section) of the EM system will be

$$\langle u_{GW} \rangle = \pi \langle S_{GW} \rangle R_0^2 = \frac{c^3 \omega^2}{32 G} (A_\otimes R_0)^2, \tag{38}$$

where $R_0$ should be the same order of magnitude at least as the spot radius $W_0$ of the Gaussian beam.

Using the first-order perturbative EM power flux $\langle u_\otimes^\phi \rangle^{(1)}_{\omega_e = \omega_g}$ and the effective power flux $\langle u_{GW} \rangle$ [see Eq.(38)] of the HFGW passing through the EM system, we can estimate "equivalent conversion efficiency" of the HFGW to the perturbative EM power flux in our EM detecting system, which is defined as

$$\eta = \frac{\langle u_\otimes^\phi \rangle^{(1)}_{\omega_e = \omega_g}}{\langle u_{GW} \rangle}. \tag{39}$$

In Table III we list the "equivalent conversion efficiencies" for the 3GHz HFGW of $h = 10^{-30}$ and the 1.3THz HFGW of $h = 10^{-28}$ in the EM detection system, the power of the background Gaussian beam is chosen as $P = 10W$.

TABLE III  The "equivalent conversion efficiencies" of the 3GHz HFGW, 1.3THz HFGW and the corresponding relevant parameters.

|     | $A$ | $\nu_g (Hz)$ | $\langle u_{GW} \rangle (W)$ | $\langle u_\otimes^\phi \rangle^{(1)}_{\omega_e = \omega_g} (W)$ | $\eta$ |
| --- | --- | --- | --- | --- | --- |
| (1) | $10^{-30}$ | $3 \times 10^9$ | $\sim 10^{-8}$ | $\sim 10^{-22}$ | $\sim 10^{-14}$ |
| (2) | $10^{-28}$ | $1.3 \times 10^{12}$ | $\sim 10$ | $\sim 10^{-20}$ | $\sim 10^{-21}$ |

Table III shows that the "equivalent conversion efficiency" of the 3GHz HFGW with $h = 10^{-30}$ will be seven orders larger than that of the 1.3THz HFGW with $h = 10^{-28}$, but the absolute value of the perturbative EM power flux $\langle u_\otimes^\phi \rangle^{(1)}_{\omega_e = \omega_g}$ produced by the latter will be two orders larger than that of the former. Recently, Fontana and R. M. L. Baker, Jr. [5] proposed a more interesting scheme (F-B scheme) in which it is possible to generate 1.3THz HFGW of $10^7 W$ by the high-temperature superconductor (HTSC) in the laboratory. If we assume further that whole such radiation power can be almost concentrated in the optimal radiative direction, and a power flux density of $10^7 Wm^{-2}$ can be made in the EM detecting system, then corresponding amplitude order of the HFGW would be $\sim 10^{-26}$. In Table IV we give the perturbative EM power fluxes and the corresponding perturbative photon fluxes generated by the 1.3THz HFGW (F-B scheme) in our EM detecting system.

TABLE IV   The perturbative parameters produced by the 1.3THz FHGW generator (Fontana-Baker scheme)

|     | $A$ | $\nu_g (Hz)$ | $P(W)$ | $\langle u_{\otimes}^{\phi(1)} \rangle_{\omega_e = \omega_g} (W)$ | $n_\phi (s^{-1})$ | $n'_\phi (s^{-1})$ |
|-----|-----|--------------|--------|-----|-----|-----|
| (1) | $10^{-26}$ | $1.3 \times 10^{12}$ | $10^5$ | $1.46 \times 10^{-16}$ | $1.69 \times 10^5$ | $9.38 \times 10^4$ |
| (2) | $10^{-26}$ | $1.3 \times 10^{12}$ | $10^3$ | $1.46 \times 10^{-17}$ | $1.69 \times 10^4$ | $9.38 \times 10^3$ |
| (3) | $10^{-26}$ | $1.3 \times 10^{12}$ | $10$ | $1.46 \times 10^{-18}$ | $1.69 \times 10^3$ | $9.38 \times 10^2$ |

In particular, since $n'_\phi$ indicates the tangential perturbative photon flux passing through the "receiving" surface $\phi = \pi/2$ ($\sim 10^{-2} m^2$) outside the spot radius $W_0$ of the Gaussian beam, and even if $P$ is reduced to 10W, we have still $n'_\phi \approx 10^3 s^{-1}$ (see scheme (3) in Table IV), our EM system should be suitable to detect the 1.3THz HFGW expected by F-B scheme.

Recently, Chiao [1-3] proposed a quantum transducers scheme with very high conversion efficiency for the HFGW and high-frequency EM waves. Under the extreme optimal condition, the conversion efficiency of a dissipationless superconductor might approach unity. Of course, since the noise and there exist unexplained residual microwave and far-infrared lost in the superconductors [1], the actual conversion efficiency would be much less than unity. In order to improve effectively the conversion efficiency, it is necessary to eliminate these microwave residual losses and to cool the superconductor down extremely low temperatures, such as millikelvins. It is unclear yet that what is an actual conversion efficiency for Chiao's quantum superconductor transducers or others. We suppose now that the power conversion efficiency of the 3GHz HFGW superconductor generators can reach up to $10^{-10}$, and assume further that one tenth of the whole radiative power can be concentrated in the effective receiver region of the EM detecting system. In this case we will be able to estimate the quantitative relation between the input EM power $P_{em}$ in the HFGW generator and the "output" power (i.e., $\langle u_\otimes^\phi \rangle^{(1)}_{\omega_e=\omega_g}$ and $n_\phi$, $n_\phi'$) in the EM detecting system (see Table V).

TABLE V  The input power in the $3GHz$ HFGW generator, the "output" power in the 3GHz HFGW EM detector and relevant parameters.

| $P_{em}(W)$ | $\eta'$ | $P_{GW}(W)$ | $u_{GW}(W)$ | $A$ | $P(W)$ | $\eta$ | $\langle u_\otimes^\phi \rangle^{(1)}_{\omega_e=\omega_g}$ (W) | $n_\phi(s^{-1})$ | $n_\phi'(s^{-1})$ |
|---|---|---|---|---|---|---|---|---|---|
| | | | | | $10^5$ | $10^{-13}$ | $0.95 \times 10^{-20}$ | $4.75 \times 10^3$ | $4.46 \times 10^3$ |
| $10^4$ | $10^{-10}$ | $10^{-6}$ | $10^{-7}$ | $10^{-30}$ | $10^3$ | $10^{-14}$ | $0.95 \times 10^{-21}$ | $4.75 \times 10^2$ | $4.46 \times 10^2$ |
| | | | | | $10$ | $10^{-15}$ | $0.95 \times 10^{-22}$ | $4.75 \times 10$ | $4.46 \times 10$ |
| | | | | | $10^5$ | $10^{-14}$ | $0.95 \times 10^{-19}$ | $4.75 \times 10^4$ | $4.46 \times 10^4$ |
| $10^6$ | $10^{-10}$ | $10^{-4}$ | $10^{-5}$ | $10^{-29}$ | $10^3$ | $10^{-15}$ | $0.95 \times 10^{-20}$ | $4.75 \times 10^3$ | $4.46 \times 10^3$ |
| | | | | | $10$ | $10^{-16}$ | $0.95 \times 10^{-21}$ | $4.75 \times 10^2$ | $4.46 \times 10^2$ |

In Table V $P_{em}$ is the total input EM power in the 3GHz HFGW generator, $\eta'$ is assumed the power conversion efficiency from $P_{em}$ to $P_{GW}$ in the superconducting generator, $u_{GW}$ is the effective HFGW power flux passing through to the EM detector, A is the dimensionless amplitude of the HFGW, $P$ is the power of the Gaussian beam, $\eta$ is the "equivalent conversion efficiency" from $u_{GW}$ to $\langle u_\otimes^\phi \rangle^{(1)}_{\omega_e=\omega_g}$ in the EM detector, $n_\phi$, $n_\phi'$ are the corresponding perturbative photon fluxes. It should be pointed out that in Table V only the power conversion efficiency $\eta'$ in the HFGW generator is assumed, while other all calculations come out of the general theory of relativity of the weak gravitational field and the electrodynamical equations in curved spacetime, and the chosen parameters are well within the current technology conditions, or they are achievable values of the present experimental possibilities.

We emphasize again that (1) since the highly directional propagated property of the perturbative photon fluxes $n_\phi$, $n_\phi'$ and the "random motion" of the remained thermal photons, requirement for the system's temperature can be greatly relaxed. (2) $n_\phi$ and $n_\phi'$ are unique nonvanishing photon fluxes passing through the surface ($\sim 10^{-2} m^2$) of $\phi = \pi/2$ or $\phi = 3\pi/2$. In this case any photon measured from such photon fluxes at the above surfaces will be a signal of the EM perturbation produced by the HFGW in the GHz to THz band. (3) If we can find a very subtle means in which the pure perturbative photon flux can be pumped out from our EM system, so that the perturbative and the background photon fluxes can be completely separated, then the possibility of detecting the HFGW would be greatly increased. More research into this subject remains to be done.

**Acknowledgements**

We are grateful to Dr. R. M. L. Baker, Jr. and Dr. P. A. Murad for their attention, helpful suggestion to this manuscript and useful remarks.

This work is supported by the National Nature Science Foundation of China under Grants No. 10175096 and the Hubei Province Key Laboratory Foundation of Gravitational and Quantum Physics under Grants No. GQ 0101.